\documentclass[letterpaper, 12pt]{article}

\usepackage{amsmath}
\usepackage{amssymb}
\usepackage{graphicx}

\title{Generalized Orthogonal Components Regression for High Dimensional Generalized Linear Models}
\author{Yanzhu Lin \and Min Zhang \and Dabao Zhang\thanks{Yanzhu Lin conducted this research as a Ph.D. student in Department of Statistics, Purdue University, West Lafayette, IN 47907; Min Zhang (email: minzhang@stat.purdue.edu) and Dabao Zhang (zhangdb@stat.purdue.edu) are associate professors in the Department of Statistics, Purdue University, West Lafayette, IN 47907. This manuscript was completed when Min Zhang and Dabao Zhang were visiting Capital Medical University, Beijing, China. This work was partially supported by NSF CAREER award IIS-0844945 and the Cancer Care Engineering project at the Oncological Science Center of Purdue University.}}

\begin{document}

\maketitle \newpage

\begin{abstract}
Here we propose an algorithm, named generalized orthogonal components regression (GOCRE), to explore the relationship between a categorical outcome and a set of massive variables. A set of orthogonal components are sequentially constructed to account for the variation of the categorical outcome, and together build up a generalized linear model (GLM). This algorithm can be considered as an extension of the partial least squares (PLS) for GLMs, but overcomes several issues of existing extensions based on iteratively reweighted least squares (IRLS). First, existing extensions construct a different set of components at each iteration and thus cannot provide a convergent set of components. Second, existing extensions are computationally intensive because of repetitively constructing a full set of components. Third, although they pursue the convergence of regression coefficients, the resultant regression coefficients may still diverge especially when building logistic regression models. GOCRE instead sequentially builds up each orthogonal component upon convergent construction, and simultaneously regresses against these orthogonal components to fit the GLM. The performance of the new method is demonstrated by both simulation studies and a real data example.
\end{abstract}


\newpage

\section{INTRODUCTION}


Available high-throughput biotechnologies have made it possible to genotype thousands of genetic markers, meanwhile, they bring challenges to statistical analyses of these data. Such data are characterized by a large number of variables ($p$) observed from a relatively small number of subjects ($n$), and create the well-known large $p$ small $n$ problems. To deal with this issue, an important strategy is to reduce the high dimensionality of the predictors before fitting models. As a supervised dimension-reduction method, partial least squares (PLS) by Wold (1975) has drawn considerable attentions, see Vinzi {\it et al.} (2010). PLS constructs orthogonal components such that these components capture information of original predictors predicting response variables, and linear models are built on the base of these components instead of the original predictors. It is computationally fast and able to take collinear or multicollinear predictors.

Success of PLS in fitting linear models motivates extensions to generalized linear models (GLMs). With the iteratively reweighted least squares (IRLS) algorithm commonly used for building regular GLMs (Green 1984), Marx (1996) proposed an extension, i.e., the iteratively reweighted partial least squares (IRPLS) algorithm, which replaces the least squares estimates with PLS estimates at each iteration. It is a natural extension of PLS, however, a different set of orthogonal components are constructed at each iteration and thus the convergence of original regression coefficients is pursued. As a result, the loadings of orthogonal components never converge, and even regression coefficients especially for logistic regressions rarely converge. A full set of distinct components at each iteration not only make it difficult to interpret, but also demand intensive computation.

Much effort has been devoted to solving the non-convergence issue of IRPLS. Ding and Gentleman (2004) applied the bias reduction procedure proposed by Firth (1993) to IRPLS, specifically for the classification problems. Firth (1993) modified the score function to remove the first order term of the asymptotic bias of maximum likelihood estimators for GLMs. Heinze and Schemper (2002) showed that this bias reduction procedure may also avoid the common infinite estimate problem of logistic regressions. However, the non-convergence issue still exists in IRPLS by Ding and Gentleman (2004), possibly due to varying components at each iteration. Alternatively, Fort and Lambert-Lacroix (2005) proposed to build up continuous pseudo-responses via ridge regression and then apply PLS to regress these pseudo-responses against the predictors; Nguyen and Rocke (2002) instead proposed to first apply PLS by treating the responses as continuous, and then fit a regular GLM using the resultant orthogonal components instead of the original predictors.



Here we propose a different strategy, namely, the generalized orthogonal components regression (GOCRE), to extend the supervised dimension reduction idea in PLS and fit high dimensional GLMs. While IRPLS repetitively constructs a different set of components at each iteration and targets a convergent set of regression coefficients, GOCRE sequentially constructs orthogonal components which maximally account for the remaining variation in the categorical outcome. The bias correction procedure by Firth (1993) is also applied. The proposed method enjoys computational privilege over IRPLS since IRPLS needs to rebuild all orthogonal components at each iteration. The construction of orthogonal components is also different from the methods by Fort and Lambert-Lacroix (2005) and Nguyen and Rocke (2002), both directly maximizing correlation between categorical responses and components.


This paper is organized as follows. The next section introduces our proposed method in details. Simulation studies are shown in Section 3, and an application of the proposed method to a real dataset is presented in Section 4. We close the paper with a brief discussion.

\section{THE METHOD} \label{Section-GOCRE}

\subsection{High Dimensional Generalized Linear Model}

Suppose the distribution of response $Y$ is a member of the exponential family distribution,
\begin{eqnarray} \label{GLM-EFD}
f(y|\theta)=\exp\left\{\frac{y \theta-b(\theta)}{a(\phi)}+c(y,\phi)\right\},
\end{eqnarray}
where $\theta$ is the canonical parameter, and $\phi$ is the known dispersion parameter. A link function $g(\cdot)$ further relates the mean of response $Y$ to the $p$ predictors in $X$, i.e.,
\begin{eqnarray} \label{GLM-Link}
g(E[Y|X])= \mu + X\beta,
\end{eqnarray}
where $\mu$ is the intercept, and $\beta$ is a $p$-dimensional column vector containing all regression coefficients of the predictors. The inverse function of $g(\cdot)$ is denoted as $g^{-1}(\cdot)$.

With a size $n$ sample $\{(y_i, \mathbf{x}_i), i=1, 2, \cdots, n\}$, a common issue is how to provide a legitimate estimate of $\beta$ in (\ref{GLM-Link}) when $p\gg n$. Denote $\mathbf{X} = (\mathbf{x}_1^t, \cdots, \mathbf{x}_n^t)^t$, an $n\times p$ matrix with rank $r_x\le min(n,p)$. The classical maximum likelihood estimators (MLEs) of $\beta$ form a space with dimension at least $p-r_x$. Suppose an $n\times r_x$ matrix $\mathbf{X}_{\cal{S}}$ is constructed by a subset of columns of $\mathbf{X}$, and further assume that there is a unique maximum likelihood estimator of $\beta_{\cal{S}}$ for the following model,
\begin{eqnarray} \label{GLM-XS}
g(E[Y|X_{\cal{S}}])= \mu + X_{\cal{S}}\beta_{\cal{S}}.
\end{eqnarray}
Correspondingly there exists a unique MLE of $\beta$, namely $\hat{\beta}$, in model (\ref{GLM-Link}), satisfying the following assumption,

\begin{description}
\item[{\it Assumption} 1.] $\hat{\beta}^t\psi=0$ whenever $\mathbf{X}\psi = \mathbf{0}_{n\times 1}$.
\end{description}

In the case that $\mathbf{X}$ is of rank $r_x$, the above assumption equivalently puts $p-r_x$ constraints on MLE $\hat{\beta}$ to make model (\ref{GLM-Link}) identifiable. This assumption makes practical sense in solving the collinearity or multicollinearity issue. For example, if the $j$-th predictor consistently doubles the value of the $k$-th predictor, we have $\hat{\beta}_j = 2\hat{\beta}_k$. Therefore, the scale of the predictor, if preserved, may indicate its importance. On the other hand, when the predictors are identical, the corresponding regression coefficients will also be identical.

Due to the aforementioned multicollinearity issue, we can focus on building model (\ref{GLM-Link}) with $\beta$ satisfying the following assumption, a population version of Assumption 1.

\begin{description}
\item[{\it Assumption} 2.] $\beta^t \psi=0$ whenever $X \psi = 0$, a.s.
\end{description}

In the next section, we consider the construction of the GOCRE model for any random pair $(Y,X)$ from the population. GOCRE sequentially builds orthogonal components $X\varpi_j$, $j=1, 2, \cdots$, to account for the variation of the categorical outcomes. For the same reason mentioned above, each $\varpi_j$ satisfies the following assumption, leading to $\beta$ satisfying Assumption 2 when a full set of components are used to build model (\ref{GLM-Link}).

\begin{description}
\item[{\it Assumption} 3.] $\varpi_j^t \psi=0$ whenever $X \psi = 0$, a.s.
\end{description}

\subsection{Generalized Orthogonal Components Regression (GOCRE)}

The orthogonal components will be sequentially constructed with a prespecified weight $w$ for each pair of $Y$ and $X$ in the whole population such that $E[w]$ is finite.  We further assume that $E[w X]=\mathbf{0}_p^t$, where $\mathbf{0}_p$ is a $p$-dimensional column vector with all components as zero. Note that such a weighted centralization of the random vector $X$ plays an important role in carrying out GOCRE. Since GOCRE constructs orthogonal components relying on a linear regression model whose response value changes at each iteration, the intercept has to be updated at each iteration (unlike PLS which removes intercept from the regression model). This weighted centralization allows separate calculation of the intercept and orthogonal components. For convenience, we denote $\nabla g^{-1}(\eta)=dg^{-1}(\eta)/d\eta$ in the following.

First, let $X_1 = X$ and for a specific $\eta$, i.e., $\eta=\eta^{(0)}$, we calculate
\begin{eqnarray} \label{Eqn-Z}
Z(\eta) = \eta + \left\{Y-g^{-1}(\eta)\right\} \big/\nabla g^{-1}(\eta).
\end{eqnarray}
A component $X_1 \alpha(\eta)$ can be constructed with $\alpha=\alpha(\eta)$ maximizing $\|E[Z(\eta) w X_1 \alpha]\|^2$ under the condition $\|\alpha\|=1$.  With a scaler variable $Z(\eta)$, we indeed have
\begin{eqnarray*}
\alpha(\eta) = E[X_1^t w Z(\eta)]/\|E[X_1^t w Z(\eta)]\|.
\end{eqnarray*}
Then regressing $Z=Z(\eta)$ against $X_1 \alpha$ with $\alpha=\alpha(\eta)$ leads to an update of $\eta$,
\begin{eqnarray}  \label{Eqn-eta}
\eta(\alpha) = E[w Z]/E[w] + X_1\alpha\gamma_1,
\end{eqnarray}
where $\gamma_1 = E[\alpha^t X_1^t w Z]/E[\alpha^tX_1^t w X_1\alpha]$. Alternatively update $\alpha(\eta)$ and $\eta(\alpha)$ until $\alpha(\eta)$ converges to $\alpha_1$, which leads to the construction of the first component $X_1 \alpha_1$.

After constructing the $j$-th component $X_j \alpha_j$, we remove $X_j \alpha_j$ from $X_j$ such that $X_{j+1} = X_j - X_j \alpha_j \theta_j$ is orthogonal to $X_j \alpha_j$, i.e.,
\begin{eqnarray} \label{Eqn-theta}
E[X_{j+1}^t w X_j \alpha_j] = 0 \Longrightarrow
\theta_j = E[\alpha_j^t X_j^t w X_j]/E[\alpha_j^t X_j^t w X_j \alpha_j].
\end{eqnarray}
Since
\[
X_j \alpha_j = X_{j-1} (I-\alpha_{j-1} \theta_{j-1}) \alpha_j  = \cdots = X \left\{\prod_{l=1}^{j-1} (I-\alpha_{j-l} \theta_{j-l})\right\} \alpha_j,
\]
we have the following preposition.

\vskip12pt %
\noindent{\it Preposition 1.} Each component $X_j \alpha_j$ can be rewritten as $X\varpi_j$ where
\[
\varpi_j = \left\{\prod_{l=1}^{j-1}(I-\alpha_{j-l}\theta_{j-l})\right\} \alpha_j.
\]
Furthermore, with the inner product defined as $\langle x,y\rangle = E[xwy]$, the components $X \varpi_1$, $X \varpi_2$, $\cdots$, are orthogonal.
\vskip12pt %

Through these first $j$ orthogonal components, we can obtain an estimate of $\eta$, say $\eta_j$. Taking $\eta=\eta_j$, we calculate $Z(\eta)$ following (\ref{Eqn-Z}). A component $X_{j+1} \alpha(\eta)$ is then constructed with
\begin{eqnarray} \label{Eqn-alpha}
\alpha(\eta) &=&\arg\max_{\alpha: \|\alpha\|=1} \{\| E[Z(\eta) w X_{j+1}\alpha] \|^2\} = E[X_{j+1}^t w Z(\eta)]/\|E[X_{j+1}^t w Z(\eta)]\|.
\end{eqnarray}
Regressing $Z=Z(\eta)$ against $X_{j+1} \alpha(\eta)$ as well as the first $j$ components leads to an update of $\eta$,
\begin{eqnarray}  \label{Eqn-etaj}
\eta(\alpha) = E[w Z]/E[w] + \sum_{k=1}^j X_k\alpha_k\gamma_k + X_{j+1}\alpha\gamma,
\end{eqnarray}
where $\gamma = E[\alpha^t X_{j+1}^t w Z]/E[\alpha^tX_{j+1}^t w X_{j+1}\alpha]$, and $\gamma_k = E[\alpha_k^t X_k^t w Z]/E[\alpha_k^tX_k^t w X_k\alpha_k]$ for $k=1, \cdots, j$. Alternatively update $\alpha(\eta)$ and $\eta(\alpha)$ until $\alpha(\eta)$ converges to $\alpha_{j+1}$, which leads to the construction of the $(j+1)$-st component $X_{j+1} \alpha_{j+1}$.

Such construction stops whenever $w^{1/2}Z(\eta)$ is uncorrelated to $w^{1/2}X_{j+1}$. Upon completion of the construction, $w^{1/2}X \varpi_1$, $w^{1/2}X \varpi_2$, $w^{1/2}X \varpi_3$, $\cdots$, are uncorrelated, which lead to the generalized orthogonal-components regression model with orthogonal components $X\varpi_1$, $X\varpi_2$, $X\varpi_3$, $\cdots$.

\vskip12pt \noindent{\it Preposition 2.} Upon completion of the construction, we can build up the generalized orthogonal-components regression model,
\begin{eqnarray}\label{Eqn-GOCRE}
g(E[Y|X])= \mu+ \sum_j \vartheta_j \left(X \varpi_j\right),
\end{eqnarray}
where $\varpi_j$, $j=1, 2, \cdots$, are as specified in Preposition 1, and $\vartheta_j$, $j=1, 2, \cdots$, are the regression coefficients of the corresponding orthogonal components. Furthermore each $\varpi_j$ satisfies Assumption 3, and $\beta=\sum_j \vartheta_j \varpi_j$ satisfies Assumption 2.
\vskip6pt

\noindent{\it Proof.} When $X_1\psi = X\psi=0$, a.s., $\alpha_1^t\psi=0$ following (\ref{Eqn-alpha}). It leads to $X_2\psi=0$, a.s.. Iteratively we have $X_j\psi=0$, a.s., and $\alpha_j^t\psi=0$. Hence $\varpi_j^t\psi=0$, $j=1, 2, \cdots$, which leads to $\beta^t\psi=0$.

\subsection{The Algorithm}

With observed data $\mathbf{Y}=(y_1, \cdots, y_n)^t$ and $\mathbf{X}=(\mathbf{x}_1^t, \cdots, \mathbf{x}_n^t)^t$, we can follow the above idea to sequentially construct orthogonal components accounting for the variation in $\mathbf{Y}$, and also provide an estimate of $\beta$ satisfying Assumption 1. The construction proceeds on the basis of prespecified weight $w_i$ for the $i$-th observation. We denote $W=diag\{w_1, \cdots, w_n\}$. Without loss of generality, we further assume that $\mathbf{X}_1 = \mathbf{X}$ has been column-wisely centralized, i.e., $\mathbf{X}^t W \mathbf{1}_n = \mathbf{0}_p$, where $\mathbf{1}_n$ is an $n$-dimensional column vector with all components as one.

Suppose that components $\mathbf{X}_1\alpha_1, \cdots, \mathbf{X}_{j-1}\alpha_{j-1}$ have been constructed, $\eta_{j-1}$ is output from the construction of the $(j-1)$-st component $\mathbf{X}_{j-1}\alpha_{j-1}$, and $\mathbf{X}_{j}$ is also constructed. We can therefore proceed to construct the $j$-th component $\mathbf{X}_j\alpha_j$, $\eta_j$, and $\mathbf{X}_{j+1}$ as follows,

\begin{itemize}
\item[\ \ ]{\bf 1.} Initialize $\eta_j = \eta_{j-1}$;
\item[\ \ ]{\bf 2.} Update $\mathbf{Z}=\eta_j + H^{-1} \{\mathbf{Y}-g^{-1}(\eta_j)\}$, with $H = diag\{\nabla g^{-1}(\eta_{j1}), \cdots,  \nabla g^{-1}(\eta_{jn})\}$;
\item[\ \ ]{\bf 3.} Update $\mu = \mathbf{1}_n^t W \mathbf{Z}/\{\mathbf{1}_n^t W\mathbf{1}_n\}$;
\item[\ \ ]{\bf 4.} Update $\alpha_j = \mathbf{X}_j^t W \mathbf{Z}/\|\mathbf{X}_j^t W \mathbf{Z}\|$;
\item[\ \ ]{\bf 5.} Update $\gamma_k = \alpha_k^t \mathbf{X}_k^t W\mathbf{Z} / \{\alpha_k^t\mathbf{X}_k^t W \mathbf{X}_k\alpha_k\}$ for $k=1, \cdots, j$;
\item[\ \ ]{\bf 6.} Update $\eta_j = \mu \mathbf{1}_n + \sum_{k=1}^{j} \mathbf{X}_k\alpha_k \gamma_k$;
\item[\ \ ]{\bf 7.} Iterate between 2-6 until $\alpha_j$ converges;
\item[\ \ ]{\bf 8.} Calculate $P_j = \alpha_j^t \mathbf{X}_j^t W \mathbf{X}_j / \{\alpha_j^t\mathbf{X}_j^t W \mathbf{X}_j\alpha_j\}$, and $\mathbf{X}_{j+1} = \mathbf{X}_j - \mathbf{X}_j\alpha_j P_j$.
\end{itemize}

Note that $\eta_{j} = (\eta_{j1}, \cdots, \eta_{jn})^t$. In Step 2, we also abuse the notations by defining $g^{-1}(\eta_j) = (g^{-1}(\eta_{j1}), \cdots, g^{-1}(\eta_{jn}))^t$.

{\bf Remark 1.} For each $k$, $\mathbf{X}_{k} = \mathbf{X}_{k-1} (I_p-\alpha_{k-1}P_{k-1}) = \cdots = \mathbf{X}\times \prod_{l=1}^{k-1} (I_p-\alpha_l P_l)$, therefore $\mathbf{X}_k^t W \mathbf{1}_n = \mathbf{0}_p$ following $\mathbf{X}^t W \mathbf{1}_n = \mathbf{0}_p$. The weighted least squares estimation equation $\mathbf{1}_n^t W\mathbf{Z} = \mathbf{1}_n^t W (\mu \mathbf{1}_n + \sum_{k=1}^j \gamma_k \mathbf{X}_k \alpha_k)$ leads to Step 3.

{\bf Remark 2.} Calculation of $P_j$ in Step 8 implies that $\mathbf{X}_{j+1}^t W \mathbf{X}_j \alpha_j = 0$. Iteratively, it leads to $\mathbf{X}_{j+1}^t W \mathbf{X}_k \alpha_k = 0$ for $k=j, j-1, \cdots, 1$. That is, the components $\mathbf{X}_1\alpha_1$, $\mathbf{X}_{2}\alpha_{2}$, $\mathbf{X}_{3}\alpha_{3}$, $\cdots$, are orthogonal when the inner product is defined as $\langle x, y \rangle = x^tWy$.

{\bf Remark 3.} Step 5 follows the application of $\mathbf{X}_k^t W \mathbf{1}_n = \mathbf{0}_p$ and $\alpha_k^t \mathbf{X}_{k}^t W \mathbf{X}_l \alpha_l = 0$, $l\ne k$, to the weighted least squares estimation equation $\alpha_k^t X_k^t W\mathbf{Z} = \alpha_k^t \mathbf{X}_k^t W (\mu \mathbf{1}_n + \sum_{l=1}^j \gamma_l \mathbf{X}_l \alpha_l)$.

{\bf Remark 4.} From the construction of the last component, say $\mathbf{X}_m\alpha_m$, we have the estimate
\begin{eqnarray*}
\hat{\beta} = \sum_{j=1}^m \left\{\prod_{k=1}^{j-1}(I_p-\alpha_kP_k)\right\}\alpha_j \gamma_j,
\end{eqnarray*}
which can be sequentially updated and satisfies Assumption 1. The parameter $\mu$ in the model (\ref{GLM-Link}) can be estimated by $\mu$ upon constructing the last component.

Compared to the original model (\ref{GLM-Link}), the GOCRE model (\ref{Eqn-GOCRE}) not only present the unique MLE satisfying Assumption 1, but also calculate the MLE without computing inverse of any matrix. While both features are desirable in analyzing $p\gg n$ data, the latter one particularly speeds up the calculation.

\subsection{Selection of Weights}

Note that, for a specific $\eta$, $Z(\eta)$ in (\ref{Eqn-Z}) has the variance
\[
var(Z(\eta)) = b''(\theta) a(\phi)\bigg/\left\{ \nabla g^{-1}(\eta)\right\}^2.
\]
Therefore, it is preferred to have a dynamic weight $w(\eta) \propto 1/var(Z(\eta))$. However, such a dynamic weight makes it impossible to construct orthogonal components in any specific inner product space.

One strategy is to take dynamic weights when the first component is being iteratively constructed with the identity matrix as the initial value. Once the first component is constructed, we have a converged weight matrix, and therefore use this weight matrix for constructing all subsequent components.

Another strategy is to run the aforementioned algorithm twice. The first run of the algorithm may take an identity weight matrix or use the above strategy to construct a weight matrix. The second run can construct a weight matrix based on the $\eta$ value from the last step of the previous run. 

Our simulation study demonstrated that the first strategy usually performs well and there is negligible gain in taking a second run of the algorithm (results not shown).

\subsection{Convergence Failures and Bias Correction}

For some GLMs, especially the logistic regression model for binary responses, MLE may not exist due to complete separation, or quasicomplete separation of different categories (Albert and Anderson 1984), and it usually results in non-convergence of the corresponding algorithm. Heinze and Schemper (2002) proposed that the penalized likelihood method by Firth (1993) can solve the convergence problem due to the aforementioned separation issue.

Suppose the model in (\ref{GLM-Link}) has log-likelihood $\ell(\mu, \beta)$ and information matrix $I(\mu, \beta)$. Instead of directly maximizing the log-likelihood function, Firth (1993) proposed to maximize the penalized log-likelihood
\[
\ell^*(\mu, \beta) = \ell(\mu, \beta) + \frac{1}{2}\log\{|I(\mu,\beta)|\},
\]
where the penalty corresponds to the Jeffreys invariant prior (Jeffreys 1946). Firth (1993) initially took this modification to reduce the bias of maximum likelihood estimates, and showed that the first order bias can be removed.

For logistic regression, we will modify our algorithm using the same idea to reduce the bias and solve the non-convergence issue. Note that only Step 2 of the algorithm in Section 2.3 needs to be modified. Define
\begin{eqnarray*}
\left\{\begin{array}{l}
\Delta=(\delta_{kl})_{n\times n}\triangleq {W}^{\frac{1}{2}}\mathbf{X}(\mathbf{X}^t W\mathbf{X})^{+}\mathbf{X}^t {W}^{\frac{1}{2}},\\
\zeta \triangleq (\delta_{11}, \cdots, \delta{nn})^t,\\
\Lambda \triangleq diag\{\delta_{11}, \cdots, \delta{nn}\},
\end{array}\right.
\end{eqnarray*}
where $(\mathbf{X}^t W\mathbf{X})^{+}$ is a Moore-Penrose pseudo-inverse. We then replace Step 2 with the following steps,
\begin{itemize}
\item[\ \ ]{\bf 2a.} Calculate $H = diag\{ (1+\delta_{11}) \nabla g^{-1}(\eta_{j1}), \cdots,  (1+\delta_{nn}) \nabla g^{-1}(\eta_{jn})\}$;
\item[\ \ ]{\bf 2b.} Update $\mathbf{Z}=\eta_j + H^{-1} \{\mathbf{Y}+\frac{1}{2}\zeta - (I_n+\Lambda)g^{-1}(\eta_j)\}$.
\end{itemize}

Note that, once the weight matrix $W$ is fixed, we then have fixed values of $\delta_{kk}$, $k=1, \cdots, n$. Therefore, unlike the IRPLS method modified by Ding and Gentleman (2004) which requires re-calculation of the weight matrix at each iteration, we do not need this re-calculation when constructing all components other than the first component, because of the fixed weight matrix $W$. Since high-dimensional data imply large matrices involved in calculation of the weight matrix, our algorithm can be more efficient in terms of computational cost.

We use GOCRE$_0$ to refer to the above implementation for bias correction. As shown in the following, the calculation of $\Delta$ can be simplified through a singular value decomposition, which will essentially speed up the computation of GOCRE$_0$.

For large $n$, it is still computationally intensive to calculate and maintain the matrix $\Delta$. Instead, Chung and Keles (2010) approximate each diagonal component of $\Delta$ with $trace\{\Delta\}/n$. A similar strategy has been utilized in constructing generalized cross-validation (Golub {\it et al.} 1979). Observing that $\Delta$ is usually of full rank and therefore $trace\{\Delta\}/n=1$ when $p\gg n$, Chung and Keles (2010) always take $\zeta = \mathbf{1}_n$ and $\Lambda = I_n$. Indeed, as shown below, $\Delta = I_n$ when $p\geq n$ and $\mathbf{X}$ is of full rank.

\vskip12pt \noindent{\it Preposition 3.} Assume $rank(\mathbf{X})=k$ and therefore the singular value decomposition $W^{1/2}\mathbf{X} = U\Omega V^t$ where $U$ is $n\times k$ with $U^t U=I_k$, $\Omega$ is a $k\times k$ diagonal matrix with positive diagonal elements, and $V$ is $p\times k$ with $V^t V = I_k$. Then $\Delta = UU^t$. Furthermore, if $k=n$, then $\Delta = I_n$; if $k=n-1$ and $\mathbf{1}_n^t W\mathbf{X} = \mathbf{0}_p^t$, then $\Delta = I_n-W^{1/2}\mathbf{1}_n\mathbf{1}_n^t W^{1/2}/\|W\|_1$, where $\|W\|_1 = \sum_{i=1}^n w_i$.
\vskip6pt

As we preprocess $\mathbf{X}$ such that $\mathbf{1}_n^t W\mathbf{X} = \mathbf{0}_{p}^t$, $\Delta$ will usually have the rank of $n-1$ instead of full rank when $p\gg n$. The above preposition implies that $\zeta = \mathbf{1}_n-W\mathbf{1}_n/\|W\|_1$ and $\Lambda = I_n-W/\|W\|_1$. Hereafter, we will use GOCRE to refer to such an implementation, that is, taking $\delta_{kk} = 1-w_k/\sum_{i=1}^n w_i$, $k=1, 2, \cdots, n$.

\section{SIMULATION STUDIES}

We simulated large $p$ small $n$ data to evaluate the performance of GOCRE and compare it with IRPLS implemented by Marx (1996) and Ding and Gentleman (2004), which are hereafter denoted by IRPLS-M and IRPLS-DG respectively. The underlying models take the logit link function in (\ref{GLM-Link}) with $\mu=0$ and $p=1000$. The predictors were divided into ten blocks, where each block was simulated from an $AR(1)$ process with the correlation $\rho$ prespecified at $\rho = 0$, $0.3$, $0.5$, and $0.7$ respectively. The regression coefficients $\{\beta_j, 1\le j\le p\}$, were generated from a Laplace distribution with location parameter two and scale parameter one.

For each different $\rho$, the simulated data consist of a training set, an independent validation set and an independent test set. Each method was used to fit the models using the training data, and the optimal number of components was chosen using the validation data. The maximum number of components is ten for all methods. The performance was evaluated based on the misclassification rate (MR) and sum of squares of the prediction residuals (PRESS) calculated from the test data. We simulated 100 data sets, each consisting of the training, validation, and test set with sample size being 100, 100, and 200, respectively.

Shown in Table~\ref{Table-SimuConvergence} are the frequencies of each method which has converged in analyzing 100 simulated data sets with each specific $\rho$. It is well known that there is a divergence problem for both IRPLS implementations (Ding and Gentleman 2004; Fort and Lambert-Lacroix 2005; Boulesteix and Strimmer 2006; Chung and Keles 2010). Indeed, the IRPLS-M did not converge in analyzing any of the simulated data sets. IRPLS-DG partially solved this problem through Firth's procedure. However, it still did not converge in analyzing, for example, 23\% of the data sets with $\rho=0.5$. On the other hand, both GOCRE$_0$ and GOCRE converged in all data analyses.

\begin{table*}[htb]
\centering \caption{Convergence Frequencies of Different Methods in Analyzing Simulated Data.}
\label{Table-SimuConvergence} \small\sf \vskip6pt
\begin{tabular}{l|c c c c}\hline\hline
Methods & $\rho=0.0$ & $\rho=0.3$ & $\rho=0.5$ & $\rho=0.7$  \\
\hline $\begin{array}{l} \text{IRPLS-M} \\ \text{IRPLS-DG} \\  \text{GOCRE}_0 \\ \text{GOCRE} \end{array}$ & %
$\begin{array}{l} 0\% \\ 79\% \\ \bf{100\%} \\\bf{100\%} \end{array}$ & %
$\begin{array}{l} 0\% \\ 82\% \\ \bf{100\%} \\\bf{100\%}\end{array}$ & %
$\begin{array}{l} 0\% \\ 77\% \\ \bf{100\%} \\ \bf{100\%} \end{array}$ & %
$\begin{array}{l} 0\% \\ 94\% \\ \bf{100\%} \\ \bf{100\%}  \end{array}$  %
\\ \hline\hline
\end{tabular}
\end{table*}

The MR and PRESS in analyzing different models are shown in Table~\ref{Table-SimuSummary} for all methods. We observe that, for each method, the higher the correlation among the predictors, the lower MR and PRESS. For either GOCRE$_0$ or GOCRE, bold MR and PRESS values indicate better performance than IRPLS-M as well as IRPLS-DG. Interestingly, IRPLS-DG reported smaller MR than IRPLS-M except for the case of $\rho=0.3$, while IRPLS-DG always reported smaller PRESS than IRPLS-M except for the case of $\rho=0.7$. In all cases, GOCRE$_0$ performed better than both IRPLS methods in terms of either criterion, except that IRPLS-M reported the smallest PRESS when $\rho=0.7$.  Indeed, GOCRE reported larger MR than IRPLS-DG only in the case of $\rho=0.0$, but performed better than both IRPLS methods in all other cases. In addition to their competitive performance and solving the convergence issue, GOCRE$_0$ and GOCRE also enjoy advantage over the other two methods in computing time which is a critical issue in analyzing high-dimensional data. As shown in the next section, both GOCRE$_0$ and GOCRE can significantly reduce the computing time.

\begin{table*}[htb]
\centering\caption{Performance Comparison in Analyzing Simulated Data. Reported are the median MR and PRESS across 100 simulated data sets, with standard errors presented in the parentheses.}
\label{Table-SimuSummary} \small\sf\vskip6pt
\begin{tabular}{c|c|cccc}\hline\hline
Criterion & Model & IRPLS-M & IRPLS-DG & GOCRE$_0$ & GOCRE \\
\hline
MR & $\rho=0.0$ & .4350(.0313) & .4250(.0349) & {\bf .4250}(.0330) & .4275(.0332) \\
& $\rho=0.3$ & .3900(.0350) & .3950(.0364) & {\bf .3825}(.0365) & {\bf .3850}(.0365) \\
& $\rho=0.5$ & .3525(.0378) & .3450(.0337) & {\bf .3350}({\bf .0336}) & {\bf .3350}({\bf .0336}) \\
& $\rho=0.7$ & .3050(.0337) & .2900(.0310) & {\bf .2850}(.0347) & {\bf .2850}(.0346) \\ \hline
PRESS & $\rho=0.0$ & .2671(.0169) & .2414(.0057) & {\bf .2405}(.0058) & {\bf .2405}(.0058)\\
& $\rho=0.3$ & .2475(.0217) & .2330(.0064) & {\bf .2313}(.0066) & {\bf .2312}(.0067)\\
& $\rho=0.5$ & .2290(.0195) & .2223(.0069) & {\bf .2208}(.0072) & {\bf .2207}(.0072)\\
& $\rho=0.7$ & .2004(.0185) & .2034(.0073) & .2034(.0084) & .2033(.0085)\\
\hline\hline
\end{tabular}
\end{table*}

As shown in Preposition 3, GOCRE$_0$ and GOCRE should report exactly the same results since we preprocessed $\mathbf{X}$ in simulated data such that $\mathbf{1}_n^{'}W\mathbf{X} = \mathbf{0}_{p}^{'}$. However, $\mathbf{1}_n^{'}W\mathbf{X} = \mathbf{0}_{p}^{'}$ could not be computationally obtained due to the computer precision. For example, our computation in MATLAB
would return centralized $\mathbf{X}$ with column means in the scale of $10^{-16}$ instead of exact zero, which resulted in the slight difference between GOCRE$_0$ and GOCRE.

\section{APPLICATION TO GENE EXPRESSION PROFILING}

Here we use the lung cancer data set (Gustafson et al., 2010) from Gene Expression Omnibus (GEO; http://www.ncbi.nlm.nih.gov/geo/) to illustrate the performance of GOCRE when compared to the two different IRPLS implementations. In this data set, a total of 187 arrays were used to monitor the expression levels of 22,215 genes from 97 cancer patients and 90 healthy individuals. For each gene, a $p$-value was obtained from the Wilcoxon rank-sum test. Ranking the genes ascendingly on the basis of their $p$-values, we constructed four data sets by including the top 1000, 2000, 5000, and all genes respectively.

\begin{figure}[htbp]

\begin{minipage}{0.5\textwidth} \centering
(a) $p=1000$
\end{minipage}
\begin{minipage}{0.5\textwidth} \centering
(b) $p=2000$
\end{minipage}
\vskip12pt

\begin{minipage}{0.5\textwidth} \centering
\includegraphics[scale=0.45]{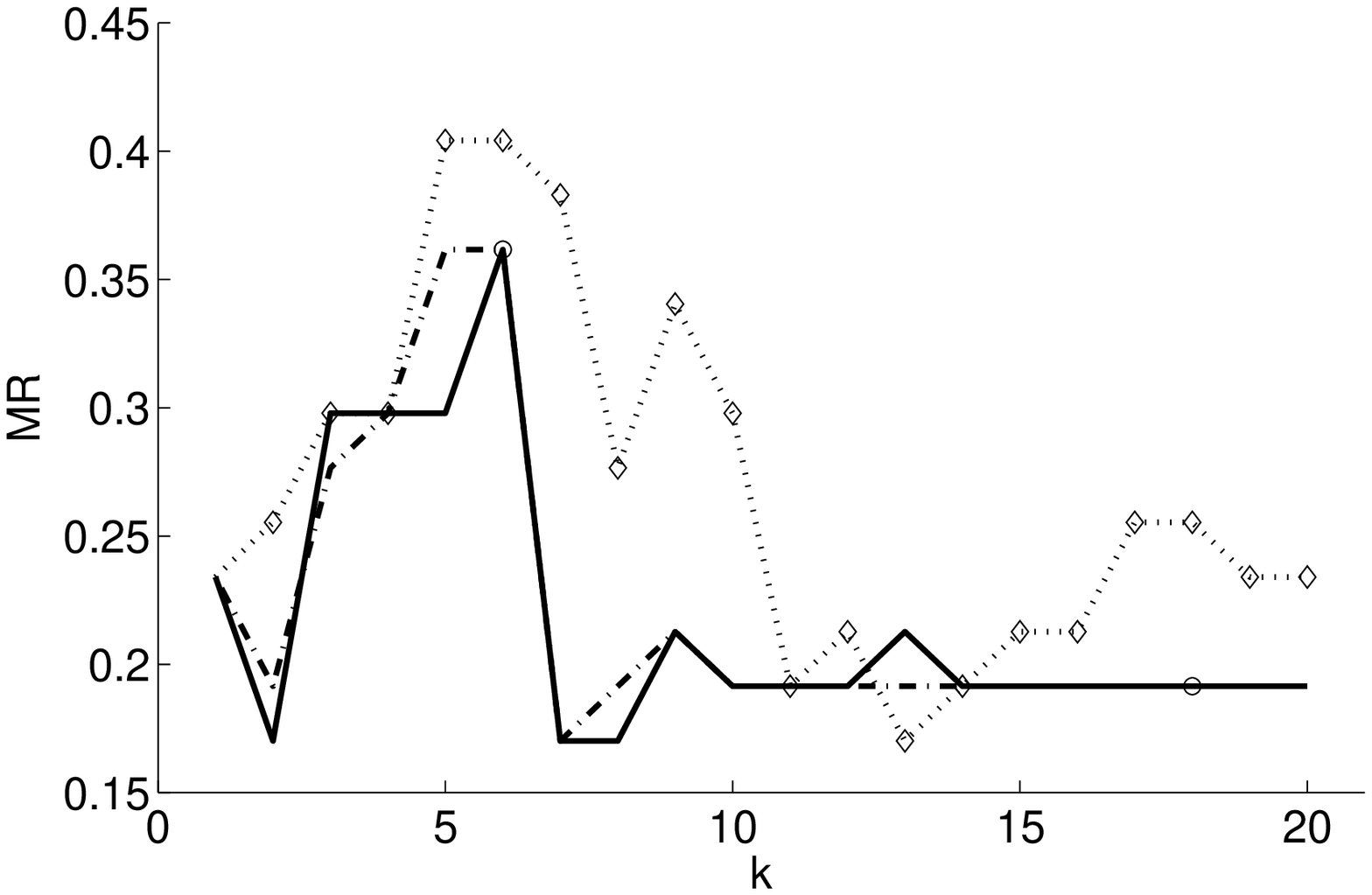}
\end{minipage}
\begin{minipage}{0.5\textwidth} \centering
\includegraphics[scale=0.45]{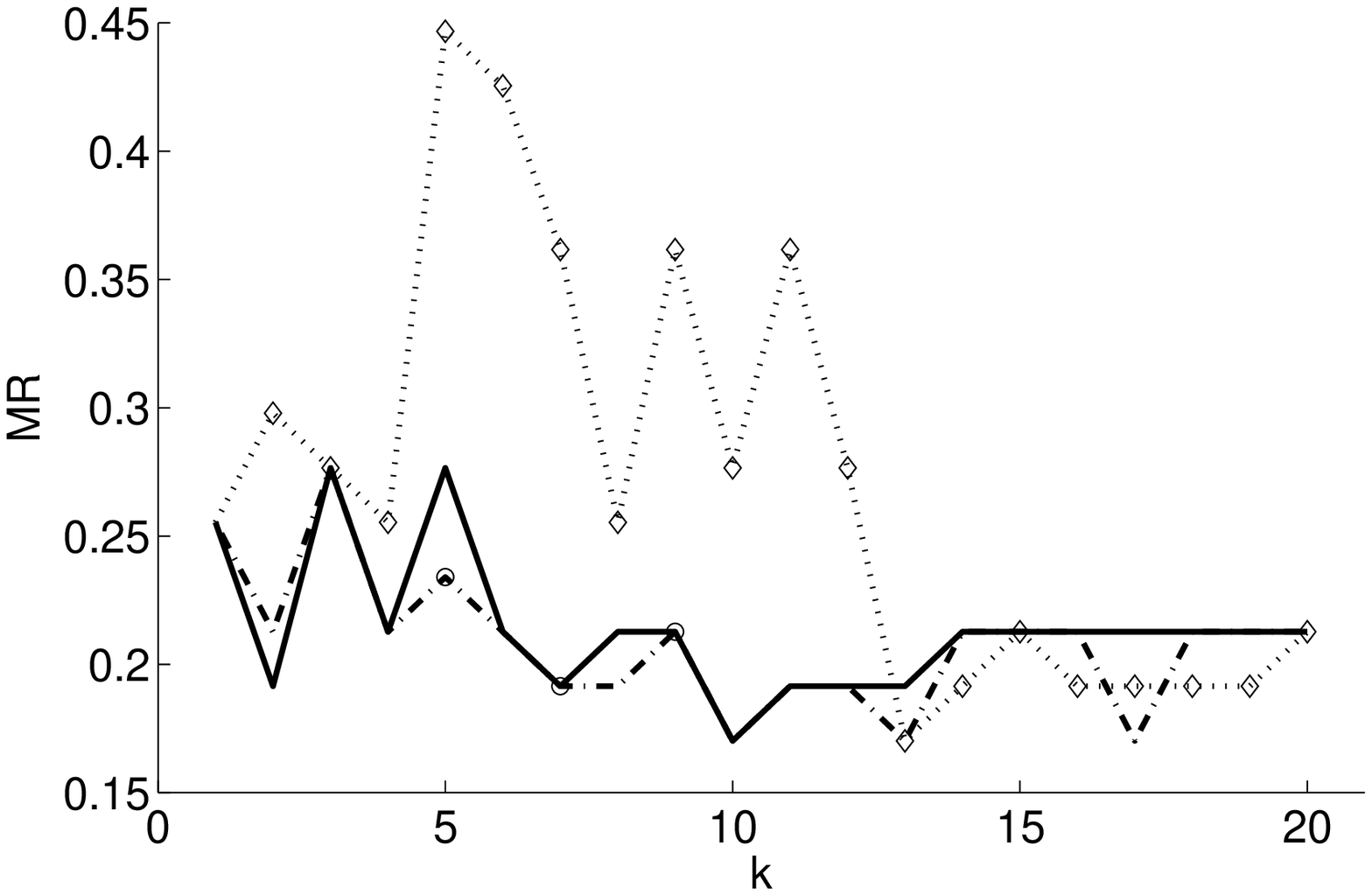}
\end{minipage}
\vskip12pt

\begin{minipage}{0.5\textwidth} \centering
(c) $p=5000$
\end{minipage}
\begin{minipage}{0.5\textwidth} \centering
(d) $p=22215$
\end{minipage}
\vskip12pt

\begin{minipage}{0.5\textwidth} \centering
\includegraphics[scale=0.45]{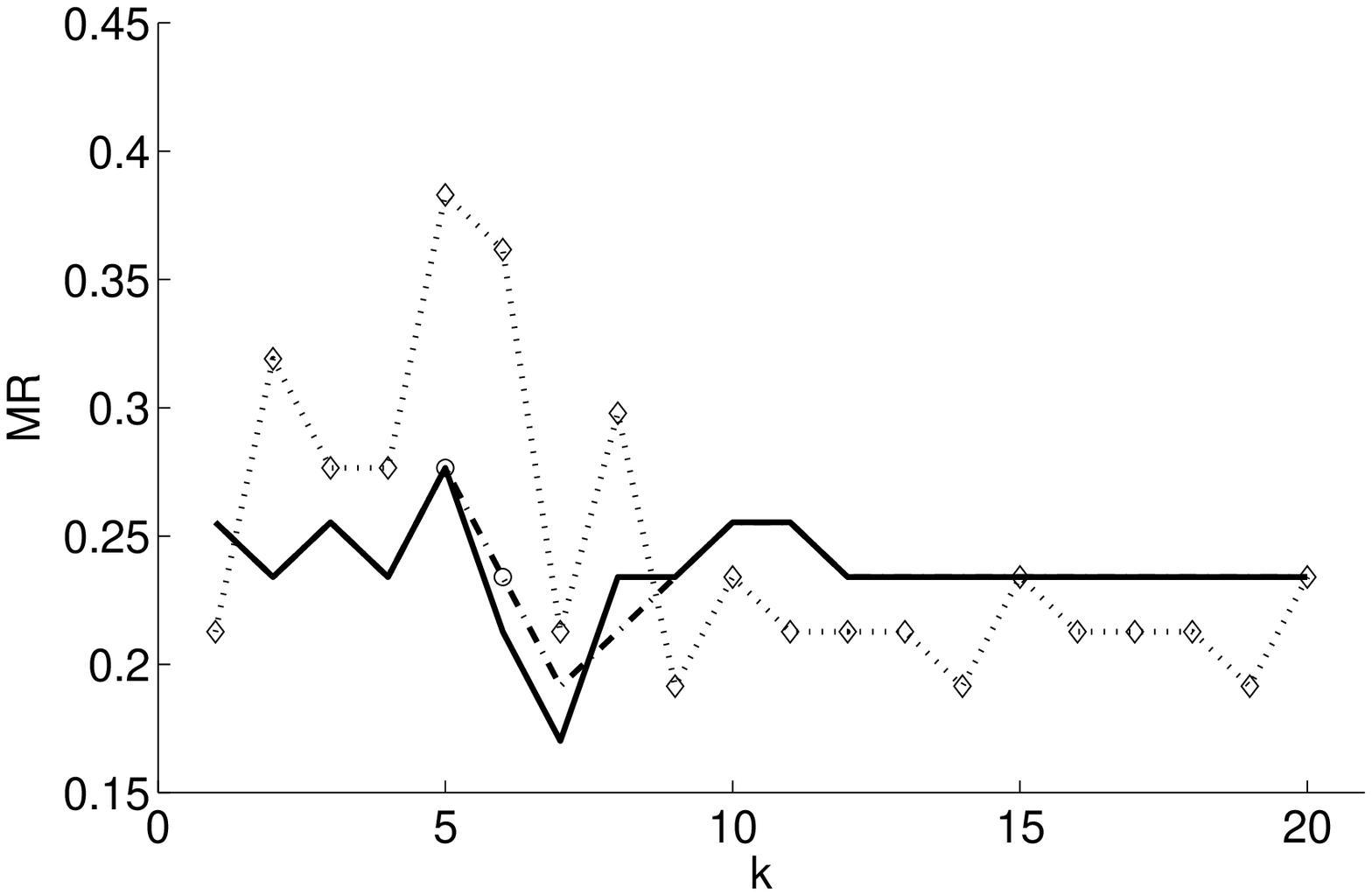}
\end{minipage}
\begin{minipage}{0.5\textwidth} \centering
\includegraphics[scale=0.45]{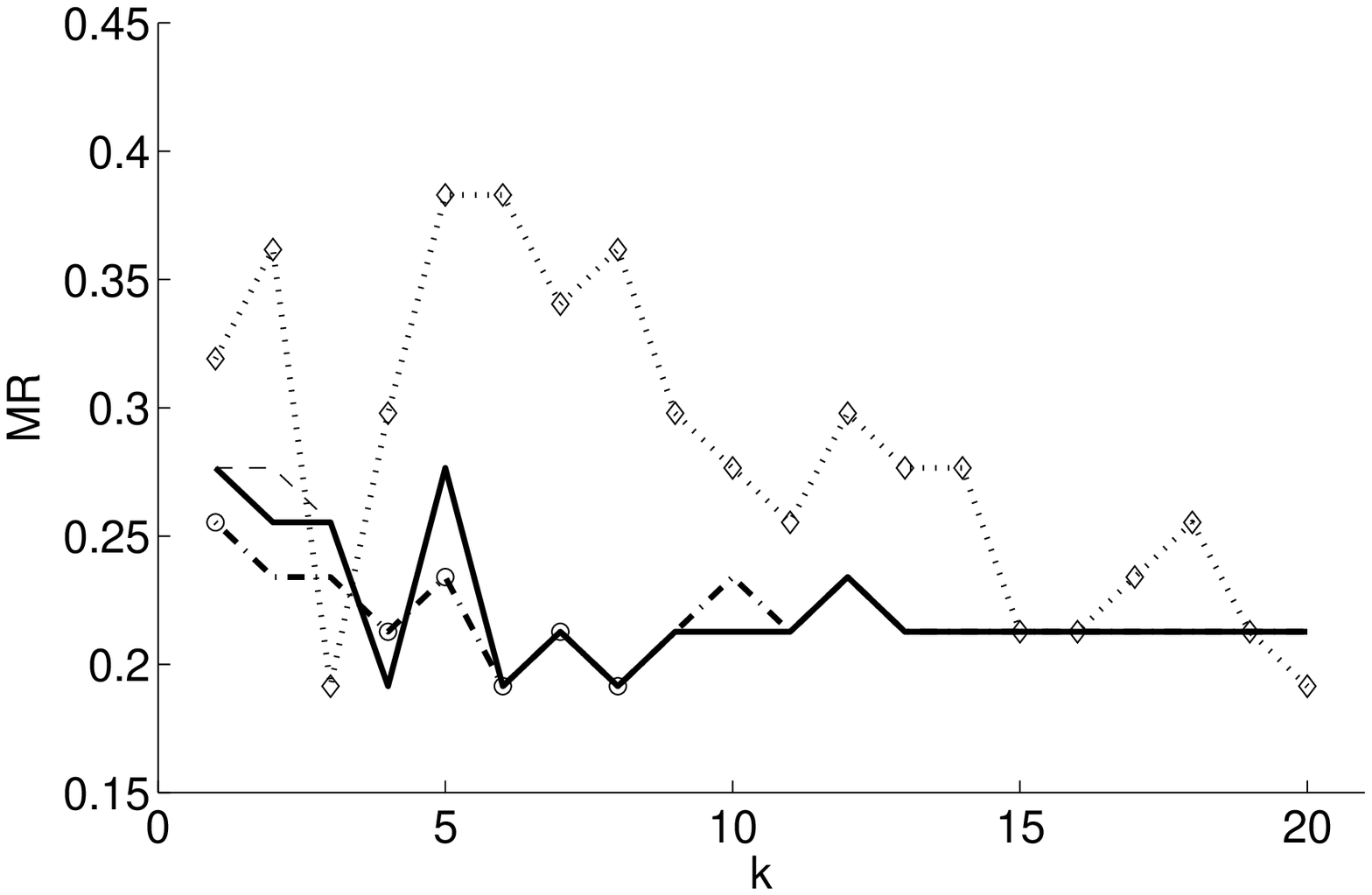}
\end{minipage}

\caption{Misclassification Rates (MR) in Analyzing the Lung Cancer Data. The results were obtained using IRPLS-M (dotted lines), IRPLS-DG (dashed dotted lines), GOCRE$_0$ (dashed lines), and GOCRE (solid lines) respectively to build up models with different number of components ($\kappa)$. Non-converged methods were marked by diamonds for IRPLS-M, and circles for IRPLS-DG.}
\label{LC-MR}
\end{figure}

\begin{figure}[htbp]

\begin{minipage}{0.5\textwidth} \centering
(a) $p=1000$
\end{minipage}
\begin{minipage}{0.5\textwidth} \centering
(b) $p=2000$
\end{minipage}
\vskip12pt

\begin{minipage}{0.5\textwidth} \centering
\includegraphics[scale=0.45]{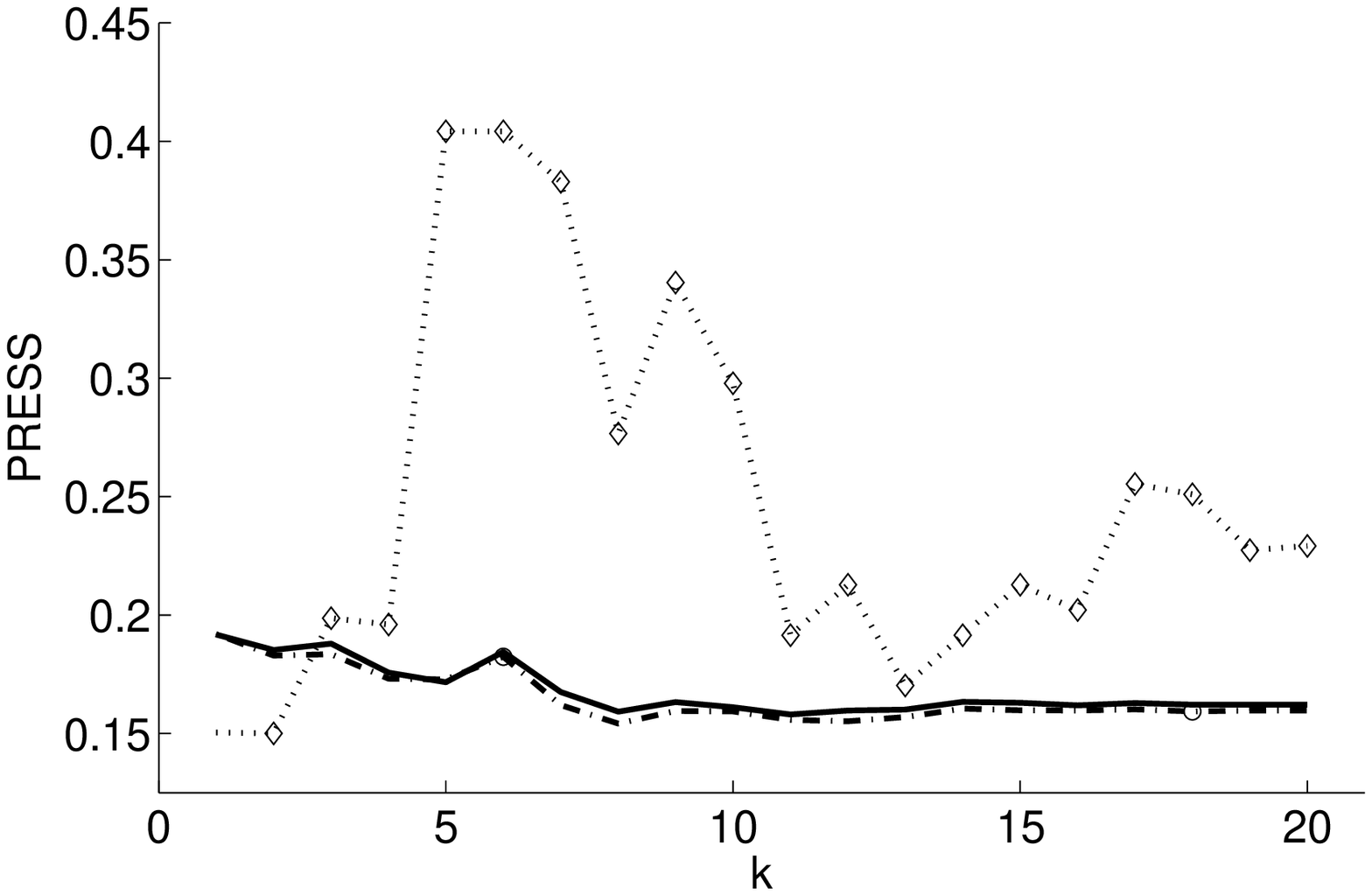}
\end{minipage}
\begin{minipage}{0.5\textwidth} \centering
\includegraphics[scale=0.45]{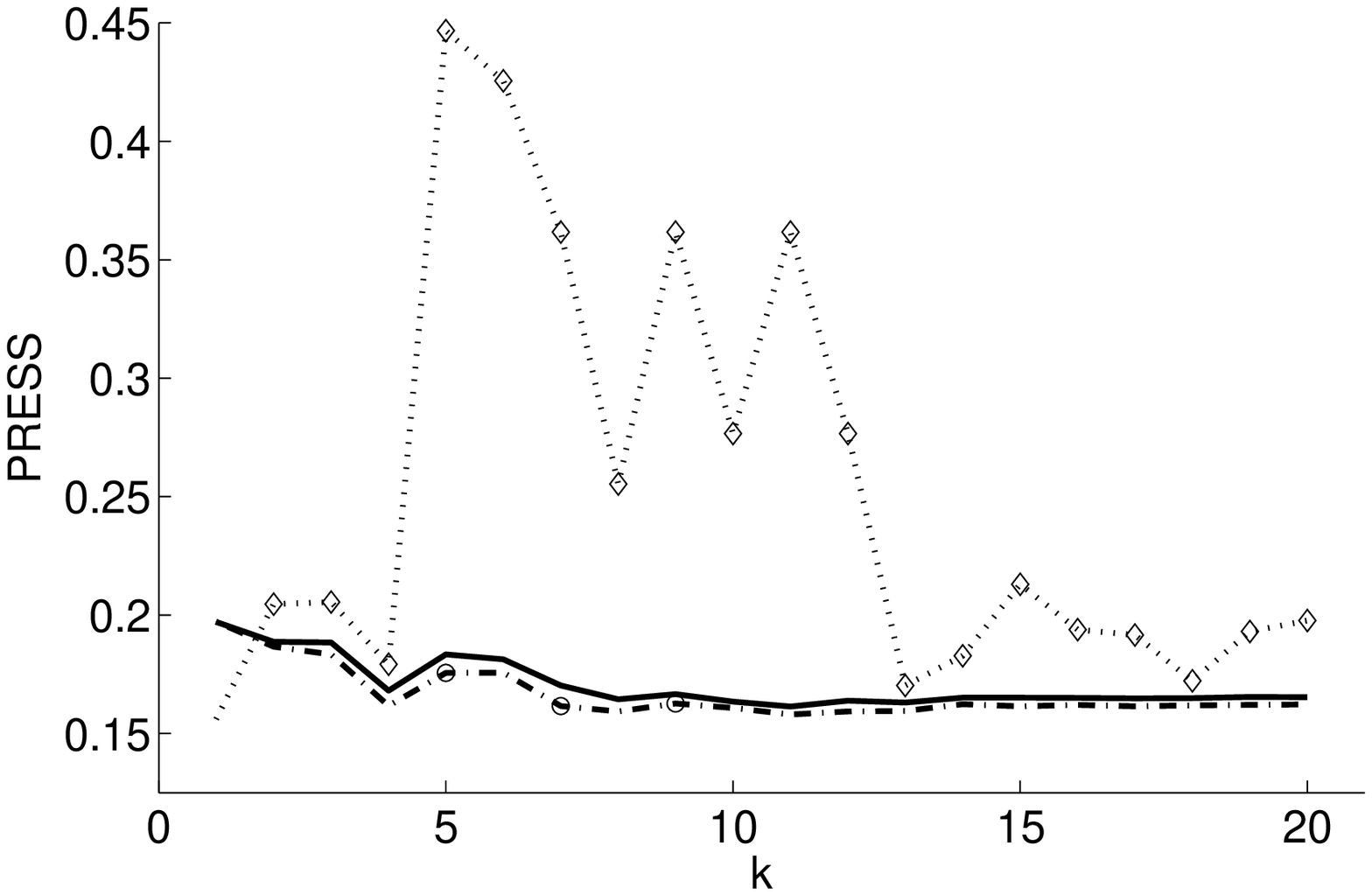}
\end{minipage}
\vskip12pt

\begin{minipage}{0.5\textwidth} \centering
(c) $p=5000$
\end{minipage}
\begin{minipage}{0.5\textwidth} \centering
(d) $p=22215$
\end{minipage}
\vskip12pt

\begin{minipage}{0.5\textwidth} \centering
\includegraphics[scale=0.45]{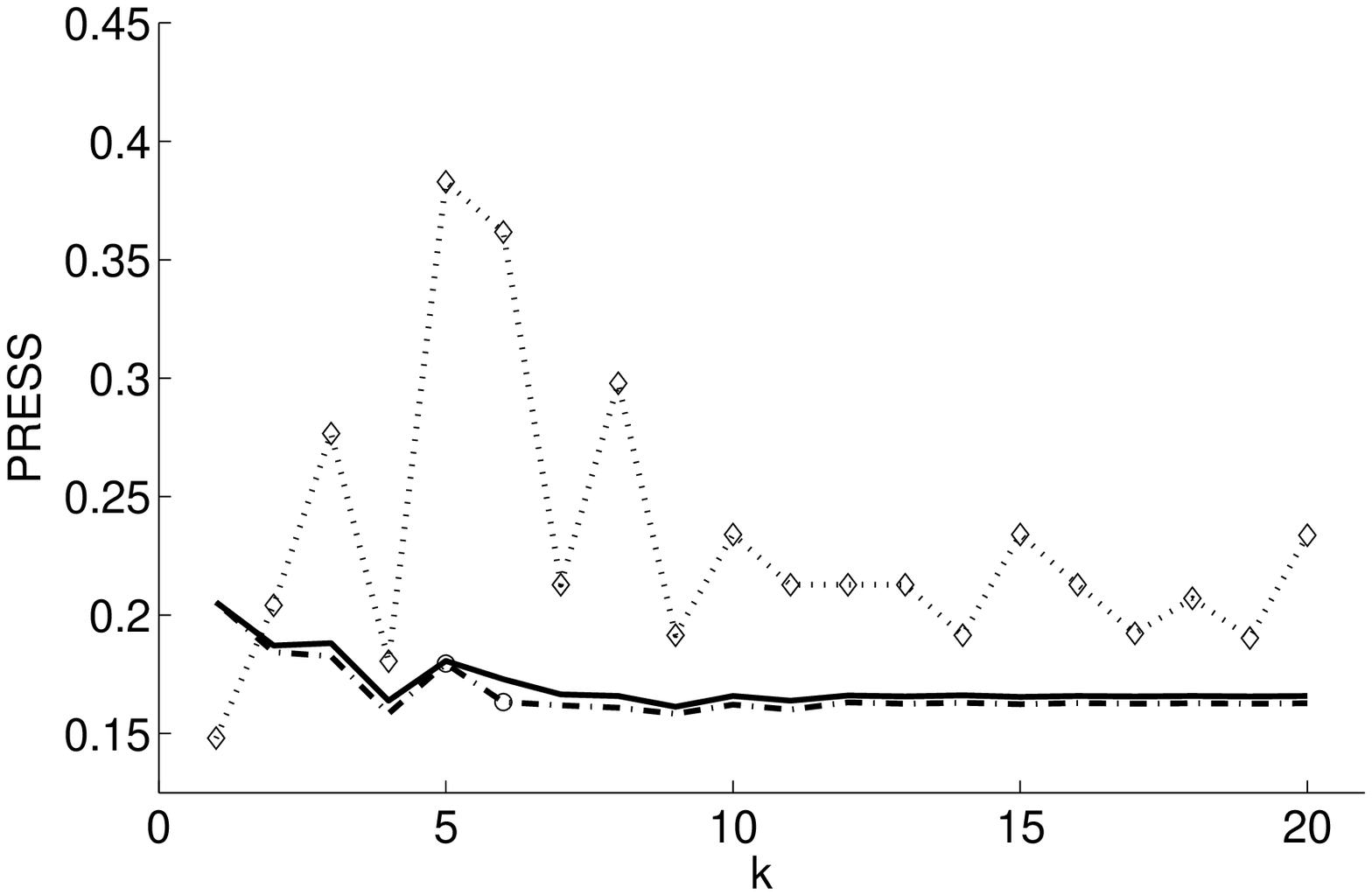}
\end{minipage}
\begin{minipage}{0.5\textwidth} \centering
\includegraphics[scale=0.45]{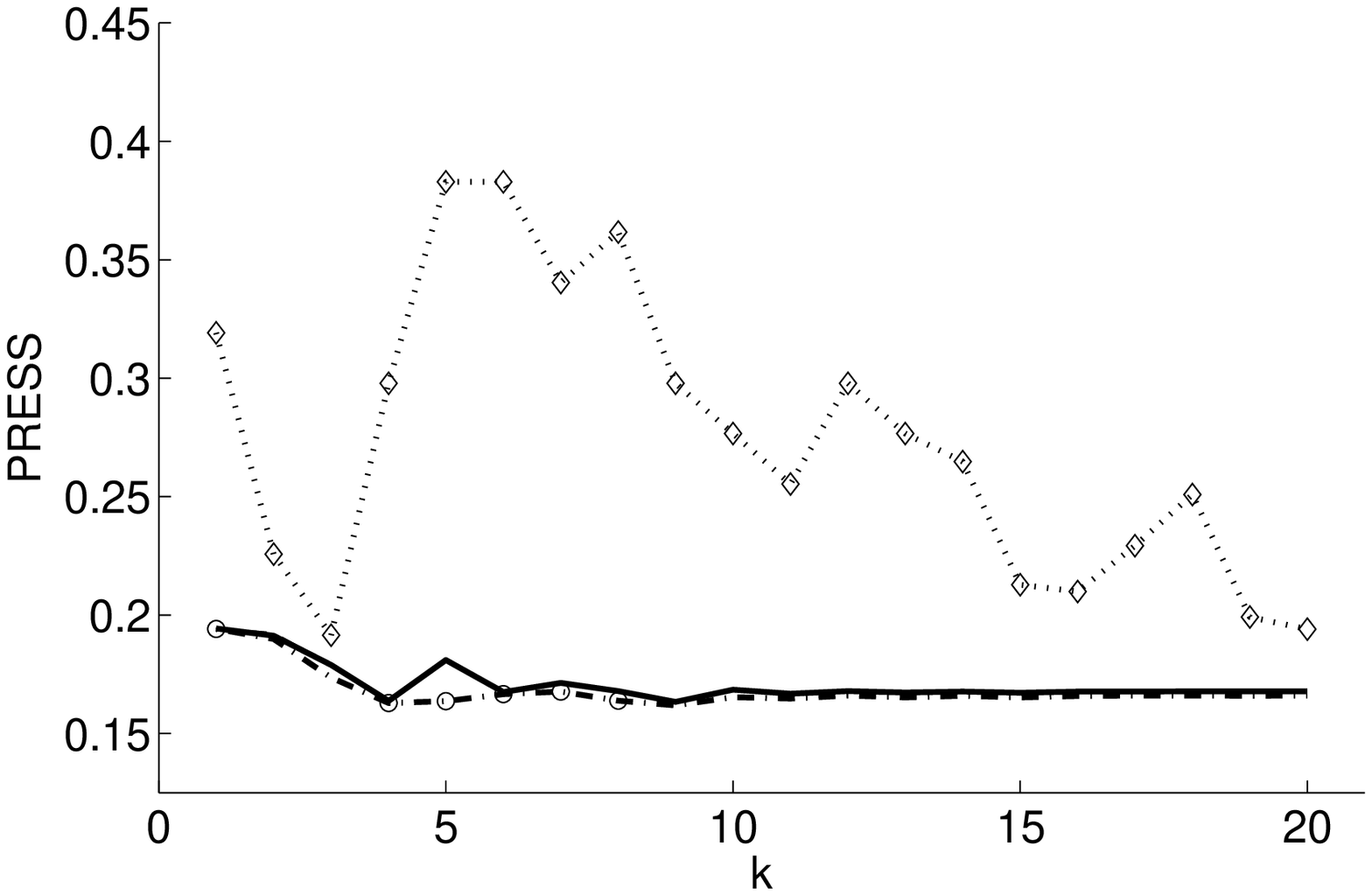}
\end{minipage}

\caption{Sum of Squares of the Prediction Residuals (PRESS) in Analyzing the Lung Cancer Data. The results were obtained using IRPLS-M (dotted lines), IRPLS-DG (dashed dotted lines), GOCRE$_0$ (dashed lines), and GOCRE (solid lines) respectively. Non-converged methods were marked by diamonds for IRPLS-M, and circles for IRPLS-DG.}
\label{LC-PRESS}
\end{figure}

For each data set, we randomly selected one-quarter of the samples to form the test data (with 24 cancer patients and 23 normal persons) and used the rest as training data. We applied each method to build up the models with different number of components (up to 20 components) using the training data, and then calculated MR and PRESS of each model based on the test data. The results are shown in Figure~\ref{LC-MR} and Figure~\ref{LC-PRESS}.

As shown in Figure~\ref{LC-MR}, IRPLS-M did not converge in constructing almost every component. Although IRPLS-DG converged in constructing majority of the components, it did not converge in constructing some components in analyzing each of the four data sets. As expected, both GOCRE$_0$ and GOCRE performed similarly, and converged in constructing each component. Except for the case with $p=5000$ where GOCRE$_0$ and GOCRE might obtain smaller MR than other methods, all methods obtained the same smallest MR in other cases.

Similar to the performance in terms of MR in Figure~\ref{LC-MR}, IRPLS-M presented rather wildly varying PRESS when different number of components were considered. The other three methods instead presented very stable and similar PRESS while the PRESS of IRPLS-DG are slightly smaller than those of GOCRE$_0$ and GOCRE. Note that the slight advantage of IRPLS-DG in PRESS did not imply improvement in MR. Indeed, all methods except IRPLS-M presented very similar MR for most models.


Not surprisingly, both GOCRE$_0$ and GOCRE were much faster than the other two methods because the IRPLS implementations need to reconstruct a set of components at each iteration, but GOCRE$_0$ and GOCRE sequentially construct all components. As shown in Table~\ref{Table-LCTime}, they took much less computing time than the other two methods when analyzing the lung cancer data using GLMs with different number of components (up to 20 components). Indeed, for each GLM model with $k$ components, both IRPLS implementations need to update the predictor matrix $k$ times within each iteration. For large $p$ small $n$ data set which produces a high dimensional predictor matrix, it is time consuming to update the predictor matrix.

\begin{table*}[htb]
\centering{\caption{Computation Time (Seconds) in Analyzing the Lung Cancer Data.}
\label{Table-LCTime} \small\sf\vskip6pt
\begin{tabular}{c|cccc}\hline\hline
p & IRPLS-M & IRPLS-DG & GOCRE$_0$ & GOCRE \\
\hline
1000 & 2,384 & 263 & 8 & 7 \\
2000 & 5,334 & 609 & 15 & 12 \\
5000 & 11,750 & 819 & 35 & 28 \\
22215 & 48,531 & 2,972 & 443 & 370 \\ \hline
\hline
\end{tabular}}
\end{table*}

\section{DISCUSSION}

Zhang {\it et al.} (2009) proposed an orthogonal-component regression (OCRE) for supervised construction of principal components which account for the variation of continuous responses. OCRE can be considered as an alternative implementation of PLS. We here propose GOCRE which extends OCRE for GLMs, focusing on binary outcomes. Such an extension makes it feasible to extend POCRE in Zhang {\it et al.} (2009) for GLMs, allowing to select variables from a large amount of candidates. However, the need of iterated procedures like IRLS for fitting classical GLMs challenges the extension. One challenge follows the use of weighted linear models to construct the components. Available components would suggest a weighted linear model with updated weights when constructing a new component. Indeed, any update of the same component in each iteration would suggest a new set of weights. Here we suggest to fix the weights upon the construction of the first component, which usually provide a satisfactory set of orthogonal components. A fixed set of weights can also be suggested using some other preliminary analysis of the data.

GOCRE is proposed to sequentially construct a set of orthogonal components and it allows to investigate the relationship between a categorical outcome and a set of variables of interest. Cross validation can be used to determine the number of orthogonal components when it is targeted to fit the underlying GLM. We may be interested in a set of orthogonal components which may take account of the variation in the categorical outcome. Therefore, the coefficient of determination as defined by Nagelkerke (1991) may be employed. An entropy measure may also be explored for such a purpose.

\section*{APPENDIX: PROOF OF PREPOSITION 3}

Expand $V$ to $\tilde{V} = (V\ V_c)$ such that $\tilde{V}^t \tilde{V} = I_p$, and further expand $U$ to $\tilde{U}$ such that $\tilde{U}^t \tilde{U} = I_p$ in the following way,
\begin{eqnarray*}
\tilde{U} = \left\{\begin{array}{ll}(U\ U_c), & if\ p\le n;\\ \left(\begin{array}{ccc} U & U_c & \mathbf{0}\\ \mathbf{0} & \mathbf{0} & I_{p-n} \end{array}\right), & if\ p > n. \end{array}\right.
\end{eqnarray*}
Accordingly, we expand $\Lambda$ to a $p\times p$ matrix $\tilde{\Lambda}$ as follows,
\begin{eqnarray*}
\tilde{\Lambda} =\left(\begin{array}{cc} \Lambda & \mathbf{0} \\ \mathbf{0} & \mathbf{0} \end{array}\right).
\end{eqnarray*}

Let $\mathbf{X}_w = W^{1/2}\mathbf{X}$ and $\tilde{\mathbf{X}}_w = \tilde{U}\tilde{\Lambda}\tilde{V}^t$, then $\mathbf{X}_w^t \mathbf{X}_w = \tilde{\mathbf{X}}_w^t \tilde{\mathbf{X}}_w = \tilde{V} \tilde{\Lambda}^2 \tilde{V}^t$, which implies that $(\mathbf{X}_w^t \mathbf{X}_w)^{+} = \tilde{V} \tilde{\Lambda}^{+2} \tilde{V}^t$, and
\begin{eqnarray*}
\tilde{\Lambda}^{+} =\left(\begin{array}{cc} \Lambda^{-1} & \mathbf{0} \\ \mathbf{0} & \mathbf{0} \end{array}\right).
\end{eqnarray*}
It follows that $\mathbf{X}_w(\mathbf{X}_w^t \mathbf{X}_w)^{+}\mathbf{X}_w^t = U\Lambda V^t \tilde{V} \tilde{\Lambda}^{+2} \tilde{V}^t V\Lambda U^t = UU^t$. That is, $\Delta = UU^t$.

When $k=n$, $U$ is an orthonormal matrix which implies that $\Delta = UU^t = I_n$.

When $k=n-1$ and $\mathbf{1}_n^t W\mathbf{X} = (W^{1/2}\mathbf{1}_n)^t \mathbf{X}_w = \mathbf{0}_p^t$, then $\tilde{U} = (U\ \ W^{1/2}\mathbf{1}_n/\sqrt{\|W\|_1})$ is an orthonormal matrix. That is, $I_n = \tilde{U}\tilde{U}^t = UU^t + W^{1/2}\mathbf{1}_n \mathbf{1}_n^t W^{1/2}/\|W\|_1$, which implies $\Delta = UU^t = I_n - W^{1/2}\mathbf{1}_n \mathbf{1}_n^t W^{1/2}/\|W\|_1$.

\section*{REFERENCES}
\begin{enumerate}
\item[] Agresti, A. (2002), {\it Categorical Data Analysis}, Wiley-Interscience, Second Edition.

\item[] Albert, A., and Anderson, J. A. (1984), ``On the Existence of Maximum Likelihood Estimates in Logistic Regression Models", {\it Biometrika}, 71, 1-10.

\item[] Boulesteix, A.-L., and Strimmer, K. (2006), ``Partial Least Squares: A Versatile Tool for the Analysis of High-Dimensional Genomic Data", {\it Briefings in Bioformatics}, 8, 32-44.

\item[] Chung, D., and Keles, S. (2010), ``Sparse Partial Least Squares Classification for High Dimensional Data", {\it Statistical Applications in Genetics and Molecular Biology}, 9, Issue 1, Article 17.

\item[] De Jong, S. (1993), ``SIMPLS: An Alternative Approach to Partial Least Squares Regression", {\it Chemometrics and Intelligent Laboratory Systems}, 18, 251-263.

\item[] Ding, B., and Gentleman, R. (2004), ``Classification Using Generalized Partial Least Squares", {\it Journal of Computational and Graphical Statistics}, 14, 280-298.

\item[] Firth, D. (1993), ``Bias Reduction of Maximum Likelihood Estimates", {\it Biometrika}, 80, 27-38.

\item[] Fort, G., and Lambert-Lacroix, S. (2005), ``Classification Using Partial Least Squares with Penalized Logistic Regression", {\it Bioinformatics}, 21, 1104-1111.

\item[] Golub, G. H., Heath, M., and Wahba, G. (1979), ``Generalized Cross-Validation as a Method for Choosing a Good Ridge Parameter", {\it Technometrics}, 21, 215-223.

\item[] Green, P. J. (1984), ``Iteratively Reweighted Least Squares for Maximum Likelihood Estimation, and Some Robust and Resistant Alternatives", {\it Journal of Royal Statistical Society}, Ser. B, 46, 149-192.

\item[] Gustafson, A. M., Soldi, R., Anderlind, C., Scholand, M. B., Qian, J., Zhang, X., Cooper, K., Walker, D., McWilliams, A., Liu, G., Szabo, E., Brody, J., Massion, P. P., Lenburg, M. E., Lam, S., Bild, A. H., Spira, A. (2010), ``Airway PI3K pathway activation is an early and reversible event in lung cancer development", {\it Science Translational Medicine}, 2, 26ra25.

\item[] Heinze, G., and Schemper, M. (2002), ``A Solution to the Problem of Separation in Logistic Regression", {\it Statistics in Medicine}, 21, 2409-2419.

\item[] Hoskuldsson, A. (1988), ``PLS Regression Methods", {\it Journal of Chemometrics}, 2, 211-228.

\item[] ------ (1992), ``The H-principle in Modelling with Applications to Chemometrics", {\it Chemometrics and Intelligent Laboratory Systems}, 14, 139-153.

\item[] Jeffreys, H. (1946), ``An Invariant Form for the Prior Probability in Estimation Problems", {\it Proceedings of the Royal Society of London, Series A, Mathematical and Physical Sciences}, 186, 453-461.

\item[] Marx, B. (1996), ``Iteratively Reweighted Partial Least Squares Estimation for Generalized Linear Regression", {\it Technometrics}, 38, 374-381.

\item[] Nagelkerke, N. J. D. (1991), ``A note on a general definition of the coefficient of determination", {\it Biometrika}, 78, 691-692.

\item[] Nguyen, D. V., and Rocke, D. M. (2002), ``Tumor Classification by Partial Least Squares Using Microarray Gene Expression Data", {\it Bioinformatics}, 18, 39-50.


\item[] Stewart, G. W. (1974), {\it Introduction to Matrix Computations}, New York: Academic Press.

\item[] Vinzi, V. E., Chin, W. W., Henseler, J., and Wang, H. (2010), {\it Handbook of Partial Least Squares: Concepts, Methods and Applications}, Berlin: Springer.

\item[] Wold, H. (1975), ``Soft Modelling by Latent Variables: The Nonlinear Iterative Partial Least Squares Approach", In {\it Perspectives in Probability and Statistics, Papers in Honour of M.~S. Bartlett}, eds J. Gani, London: Academic Press.


\item[] Zhang, D., Lin, Y., and Zhang, M. (2009), ``Penalized Orthogonal-Components Regression for Large p Small n Data", {\it Electronic Journal of Statistics}, 3, 781-796.
\end{enumerate}

\end{document}